# The Potential of Planets Orbiting Red Dwarf Stars to Support Oxygenic Photosynthesis and Complex Life

Joseph Gale[1] and Amri Wandel[2]

*The Institute of Life Sciences[1] and The Racach Institute of Physics[2], The Hebrew University of Jerusalem, 91904, Israel.*
*gale.joe@mail.huji.ac.il[1] amri@huji.ac.il[2]*

## Abstract

We review and reassess the latest findings on the existence of extra-solar system planets and their potential of having environmental conditions that could support Earth-like life. Within the last two decades, the multi-millennial question of the existence of extra-Solar-system planets has been resolved with the discovery of numerous planets orbiting nearby stars, many of which are Earth-sized (and some even with moderate surface temperatures).

Of particular interest in our search for clement conditions for extra-solar system life are planets orbiting Red Dwarf (RD) stars, the most numerous stellar type in the Milky Way galaxy. We show that including RDs as potential life supporting host stars could increase the probability of finding biotic planets by a factor of up to a thousand, and reduce the estimate of the distance to our nearest biotic neighbor by up to 10. We argue that multiple star systems need to be taken into account when discussing habitability and the abundance of biotic exoplanets, in particular binaries of which one or both members are RDs. Early considerations indicated that conditions on RD planets would be inimical to life, as their Habitable Zones (where liquid water could exist) would be so close as to make planets tidally locked to their star. This was thought to cause an erratic climate and expose life forms to flares of ionizing electro-magnetic radiation and charged particles. Moreover, it has been argued that the lesser photon energy of the radiation of the relatively cool RDs would not suffice for Oxygenic Photosynthesis (OP) and other related energy expending reactions. Recent calculations show that these negative factors are less severe than originally thought.

Numerous authors suggest that OP on RD planets may evolve to utilize photons in the infrared. We however argue, by analogy to the evolution of OP and the environmental physiology and distribution of land-based vegetation on Earth, that the evolutionary pressure to utilize infrared radiation would be small. This is because vegetation on RD planets could enjoy continuous illumination of moderate intensity, containing a significant component of photosynthetic 400-700 nm radiation.

On Earth, OP has been an essential factor in producing the biosphere environment, which enabled the appearance and evolution of complex life. We conclude that conditions for OP could exist on RD planets and consequently the evolution of complex life (as we know it) might be possible. Furthermore, the huge number and the long lifetime of RDs make it more likely to find planets with photosynthesis and life around RDs than around solar type stars.

**Key words:** Astrobiology; Bio-signatures; Extra-solar terrestrial planets; Habitability; Oxygenic-Photosynthesis; Red-Dwarf stars.

# 1. Outline

In this paper, we will show that planets of Red Dwarf stars (RDPs) are more hospitable to the evolution of biotic life ("habitable") than thought before, and demonstrate the implications of the habitability of RDPs on the expected abundance of biotic planets.

In the first ("astronomy") sections of this paper, we show how the results of the Kepler space telescope have revolutionized our knowledge of the statistics of extra-solar system planets (exoplanets). We focus especially on Red Dwarfs, by far the most numerous stellar population in the Milky Way. We review the recent work showing how conditions on RDPs may be more hospitable to life than thought earlier. In particular we show that conditions favorable to life may exist on tidally locked RDPs, as well as in multiple star systems, and that planetary water and Oxygen can survive the relatively strong UV and X-ray outbursts during the early flare-phase in the evolution of Red Dwarfs.

In the subsequent ("biology") sections, we consider the suitability of RDPs to Oxygenic Photosynthesis (OP), the process used by plants and other organisms to convert light into chemical energy. On Earth, OP uses radiation in the 400-700 nm waveband, which is considered to have been a prime requirement for the evolution of complex, multi-cellular life. On RDPs, Near Infrared radiation (NIR, >700 nm) is by far more copious than on planets orbiting solar type stars. Many authors have thus described mechanisms whereby photosynthesis could evolve to utilize NIR. We argue that, based on analogy to the evolution of photosynthesis and to the distribution of vegetation on Earth, the visible radiation (<700 nm) incident on RDPs, may be more than adequate to support OP (Gale and Wandel 2015). Finally, we argue that the product of the huge number of RDPs and their ability to support OP significantly enhances the statistical chance for life supporting planets in our Milky Way galaxy.

In Section 2 we review the most recent Kepler and other exoplanet data, in particular the implications for the existence of habitable Earth-sized planets of RDs. In Section 3 we review the general conditions on RDPs that may allow Earth-like life and in Section 4 we discuss the habitability of multi-star (in particular RD) systems. The implications on habitability of the energetic flux in the early evolutionary stages of RD stars are given in Section 5. Section 6 examines the implications of tidal locking to the presence of liquid water and to habitability of RDPs. In Section 7, we survey how OP has been vital for the evolution of complex life on Earth. In Section 8, we discuss the environment on RDPs in relation to OP and plant growth, and make the analogy to certain regions on Earth. Finally, in Section 9, we demonstrate by a quantitative calculation, how including RDPs enhances the abundance of putative biotic planets, presumably suitable for Earthlike life.

## 2. Prevalence of Earthlike planets outside of the Solar System

In the past decade planets have been discovered around hundreds of nearby stars, yet most of them were Jupiter-like gas giants, and too close to their host star to permit liquid water on their surface (e.g. Fridlund *et al*. 2010). In the last four years, the Kepler mission yielded over 4000 exoplanet candidates, most of them with sizes smaller than Neptune and down to Earth-sized planets. Such small planets (Super Earths) have been shown to be abundant and even to constitute the majority of exoplanets (Buchhave *et al*., 2012; Batalha *et al*., 2013). Further work has demonstrated that such planets are often found within the Habitable Zone (HZ) of their host star. Recent analyses of the Kepler data showed (Petigura *et al.,* 2013) that about 20% of all solar-type stars have small, approximately Earth-sized planets orbiting within their HZ. Observational uncertainties and false-positive detections (Foreman-Mackey, Hogg and Morton, 2014; Farr, Mandel and Stroud, 2014) may significantly reduce this figure (down to 2-4%, however with a large uncertainty), yet it still implies a significant fraction and a huge number of stars with Earth-size habitable planets. Similar results have been obtained by different methods. The HARPS team (using the Doppler method) estimated that more than 50% of solar-type stars harbor at least one planet, with the mass distribution increasing toward the lower mass end (<15 Earth masses) (Mayor *et al*., 2011). HARPS also detected Super Earths in the Habitable Zone (Lo Curto *et al*., 2013). These findings demonstrate that "Earthlike" planets (in the sense of Earth-size planets in the HZ) are probably quite common, enhancing the probability of finding planets with conditions appropriate for the evolution of biological life as we know it.

Another conceptual break-through has been the recognition that life need not be limited to planets orbiting solar-type stars, which are only a small fraction of all stars in the Galaxy. Especially Red Dwarf (RD) or M-type stars (the lowest-mass stars, 0.08-0.6 $M_{Sun}$, less luminous and cooler than the Sun), are of great interest, as RDs are the type most common in the Milky Way galaxy, constituting about 80% of the stars in our neighborhood. In this review, we focus our discussion on planets of RD stars (RDPs).

The Kepler results show that a significant fraction of all RDs may have Earth-sized planets (Dressing and Charbonneau, 2013; 2015). These authors find that RDs have on average $\{0.56\}_{-0.05}^{+0.06}$ Earth-size planets (1-1.5 $R_{\oplus}$) and $\{0.46\}_{-0.05}^{+0.07}$ super-Earths (1.5-2 $R_{\oplus}$) with periods shorter than 50 days. Recent estimates of the occurrence of Earth sized planets in the HZs around Kepler RDs give that over 40% are expected to harbor an Earth-sized planet in their HZ (Kopparapu, 2013); the more recent work of Dressing and Charbonneau (2015) estimates that within a conservatively defined HZ, based on the moist greenhouse inner limit and maximum greenhouse outer limit, RDs have on average $\{0.16\}_{-0.07}^{+0.17}$ Earth-size planets and $\{0.12\}_{-0.05}^{+0.10}$ super-Earths. Adopting broader HZ boundaries (Venus to Mars) yields a higher estimate of $\{0.24\}_{-0.08}^{+0.18}$ Earth-size planets and $\{0.21\}_{-0.06}^{+0.11}$ super-Earths per RD. If correct, this would mean that nearly every second star in the Milky Way has a terrestrial planet within its Habitable Zone and that the nearest potentially habitable Earth-size planets could be less than ten light years away (<3pc; as this review is intended to be interdisciplinary, we will use light years rather than parsecs which are more common in astronomy publications) . If most Earthsized HZ RDPs were suitable for life (as we will discuss in the following sections) this would imply that about half of the stars in the Milky Way could harbor a biotic planet. Given these considerations, one may calculate the likely abundance of life using the analyses of Wandel (2015).

## 3. Potential clement conditions for life on RD planets

Despite the great advances of the last fifteen years in the detection of extra-solar-system planets, we still have no evidence for the existence of life anywhere but on Earth. Consequently, discussions of life and the ecological potential for life are constrained by the caveat of "life as we know it". This will also be the limitation of the following discussion, although we recognize that life elsewhere may be very different and have environmental-ecological requirements dissimilar to those with which we are familiar (Pohorille and Pratt, 2012; Azuo-Bustos and Vega-Martinez, 2013). In this review, "clement conditions" refer especially to the requirements of OP.

Much single celled life exists below the surface of Earth, where it utilizes sources of energy other than light (Gold, 1999). Some complex life has developed on Earth in the dark, in ecosystems originating around sub-ocean hydrothermal vents. These ecosystems are presently exceptional, although very significant, as they may have been the source of Earth's first life (van Dover, 2000). Furthermore, the existence of such hydrothermal life, which utilizes, non- photosynthesis sources of energy, is of great interest for the search for life in ecological niches outside of Earth. This could be either within the Solar system, such as in the subsurface liquid water of the satellites Europa (of Jupiter) and Enceladus (of Saturn), or outside the Solar system, on planets (and their satellites) of other stars.

Early photosynthesis on Earth did not split water and release oxygen (Blankenship, 1992, Hohmann-Marriot and Blankenship, 2011). Many examples of such photosynthesis are still extant (e.g. Ritchie and Runcie, 2013). Although today much life on Earth is not dependent on OP, nearly all advanced multi-cellular life is: either directly (plants) or indirectly (herbivores, carnivores and parasites). While transmuting radiant to chemical energy and providing food, OP also supports the oxygen rich atmosphere essential for the aerobic respiration that drives complex life forms and provides protection from excessive UV radiation (Wayne, 1992; and see below). The possibility and conditions for life on RDPs, have been discussed by many authors (e.g. Heath, *et al.*, 1999; Tarter *et al.*, 2007), long before the discoveries of the Kepler telescope. In the next three sections, we briefly summarize these major "astrophysical" considerations, which show that the basic ingredients of life, water and oxygen, can persist on RDPs. We then argue that conditions favoring OP may exist on RDPs and extend this argument, with reference to conditions in certain analogous ecosystems on Earth.

# 4. Habitability of planets in a multi-star system

Most stars are part of a binary or multiple system (65-100% in the middle of the Main Sequence, e.g. Duquennoy and Mayor 1991, Duchene and Kraus 2013). For low mass stars such as Red Dwarfs the fraction may be somewhat lower (e.g. Bardalez et al., 2015, find ~45%, but they admit that they may have failed to identify some of the objects in their sample as binary or multiple systems). Even if only half of the apparent RDs were actually binary systems, the fraction of individual RDs in such systems would be ~67%. For the purpose of estimating the abundance of habitable planets, it is therefore important to know whether plantes of binary or multiple stars are habitable or not.
Early researches suggested that planets in such systems would have highly erratic orbits. This would result in chaotic climates – most unsuitable for the evolution of advanced life forms. However, recent calculations show that planetary orbits in multiple star systems could be stable, producing climates, albeit different from those on Earth, but not impossible for life (Joshi, 2003). Thus, multiple star systems need to be taken into account when discussing habitability and the abundance of biotic exoplanets. Here we show that this argument is even stronger for RD stars.

Planets orbiting much closer to one of the stars of a binary (or multiple) star system than the distance between the stars are dominated by the radiation from the nearby star. Hence their climate is relatively weakly influenced by the radiation from the other star(s) in the system (unless the other star(s) is much brighter than the planet's host star). This particular situation is especially likely for RDPs for two reasons. First, binary systems consisting of two RDs are by far the most common of all multiple star systems. Second, the Habitable Zones of RD stars are significantly smaller than those of solar-type (or other) stars - of the order of ~0.1-0.4 AU, and hence more likely to be smaller than the distances between stars in a binary or multiple system.

In unfavorable orbital configurations, planets in multiple star systems may be dynamically unstable (the planet would be expelled from its orbit relatively quickly, being either ejected from the system altogether or moved to a closer or further-out orbit). Even orbits that are stable may be unsuitable for the evolution of life because of extreme variations in surface temperature during different parts of the orbit. As described above, planets that orbit just one star in a binary pair (a configuration called "S-type" orbit), may have a relatively isothermal biosphere if the Habitable Zone is much smaller than the distance between the two stars of the binary, provided the other star is not much brighter.

In general, the condition for an isothermal planet in an S-type orbit is

$$D_p << D_{12} \ (L_2/L_1)^{1/2} \qquad \text{(eq. 1)}$$

where $D_p$ is the distance of the planet from the host star, $L_1$ is the luminosity of that star, $D_{12}$ is the average distance between the two components of the binary and $L_2$ is the luminosity of the other star in the binary system. For planets orbiting around both stars ("P-type" or "circum-binary" orbits) the condition for thermal stability is obviously $D_p >> D_{12}$, that is, the planet should be very far from the host binary, so it is weakly affected by the changing positions of the two stars relative to the planet.

It is estimated that 50–60% of all binary stars are capable of supporting habitable terrestrial planets within stable orbital ranges. This figure comes from simulations demonstrating that the presence of a binary companion may improve the rate of planet formation with stable orbits by stirring and increasing the accretion rate of the protoplanets within the proto-planetary disk (Quintana and Lissauer, 2010). As argued by these authors, detecting planets in multiple star systems is more difficult than around single stars, whether by the Doppler or the transit method, which may explain the relatively low numbers of planets actually observed in binary or triple star systems. However, the lower rate of planets around multiple stars may be real, as demonstrated by recent analyses of the Kepler database (corrected for observational bias). This could imply that a stellar companion at distances smaller than 1500 AU suppresses planet formation (Wang et.al. 2014). This observation is supported also by the indirect evidence in close binary systems with RDs, that a close companion my disrupt the proto planetary disc and thus limit or stop planet formation (Meibom Mathieu and Stassun, 2007). On the other hand in certain extended (>1000 AU) stellar systems planetary systems appear to be as common as around single stars (Zuckerman 2014). In conclusion, it presently seems that multiple stellar systems may have habitable planets and hence should be taken into account for estimating the abundance of biotic planets. This conclusion is much stronger if one or more of the stars of a binary (or multiple) system is a RD star. Not only most of the stars in the Milky Way are found in such binaries, but planets in S-type orbits around a RD-member of a binary are more likely to have a stable climate, than those around a solar-type (or other) member of a multiple-star system.

## 5. Energetic radiative flux at early evolutionary stages of RDs

Young RDs have strong coronal X-ray radiation with values of X-ray-to-total luminosity ratio up to 100 times that of solar-type stars, and exhibit strong coronal UV radiation flares and UV surface fluxes up to 100 times those of comparable age (or rotation) solar type stars (e. g. Houdebine 2003). The energy of these flares may be estimated by measuring their UV lines, in particular the hydrogen Lyman-$\alpha$ line which has been shown to correlate with the UV emission of RDs (Shkolnik *et al.* 2014; Shkolnik and Braman, 2014; Jones and West 2015).

Rugheimer et al. (2015) recently examined how UV emission affects Earth-like planets orbiting M dwarfs, in particular the detectability of their life signatures. They model the atmospheres and spectra of Earth-like planets orbiting RDs with and without UV radiation, as well as observed RDs with UV radiation data. They focus on the effect of UV activity levels on detectable atmospheric features that indicate habitability on Earth, such as $H_2O$, $O_3$, $CH_4$, $N_2O$, and $CH_3Cl$.

Taking into account that planets within the Habitable Zone of RDs are much closer to the host star (0.1-0.4 AU) than in the case of solar type stars (~1 AU), these radiation events could be dangerous for both, the atmosphere and surface life. For example, it has been thought that the strong XUV activity of RDs could strip the atmosphere off RDPs (Tian 2009). Indeed, high UV radiation and charged particle flux may cause an evaporation of the planet's atmosphere, which could explain the high fraction of hot dense planets around cool stars in the current exoplanet population (e.g., Wu and Lithwick 2013). This led many astronomers and astrobiologists to assume that life (as we know it) cannot evolve on RDPs.

However, in recent years the possibility for the evolution of life on planets around RDs has been reassessed (e.g. Tarter *et al.*, 2007). Guinan and Engle (2013) found that during the first 3 billion years of the life of a RD, its X-ray radiation level decreases by a factor of 30, bringing it down to the level of solar type stars (relative to the optical luminosity). Moreover, the lifetime of RDs is much longer than that of solar type stars. Calculations show that, depending on their mass, RDs could live over 100 billion years, much longer than the (current) age of the universe. Since the huge radiation flares happen only in the early evolution of RDs (e.g. Silvestri, Hawley and Oswalt 2005, West *et al.* 2008), this leaves a long period during which the level of coronal X-rays and radiation flares of RDs are low. Life on Earth took less than 4 billion years to evolve from the first mono-cellular life to multi-cellular complex life. If this is a characteristic duration, it seems that RDPs would experience a long enough low-radiation epoch to allow the evolution of life.

Also the atmospheric-stripping effect of the XUV radiation during the early epoch of RDs has been reassessed (Erakev et al. 2013). Recent calculations demonstrated that RDP atmospheres may survive the high XUV radiation epoch. For example, a super-Earth planet with 6–10 Earth masses may retain a primary $CO_2$ atmosphere even while orbiting a young RD with XUV radiation levels of up to 1000 that of the Sun (Tian, 2009). The latter may be present during the early evolutionary phases of RDs. Atmospheres consisting of lighter gases (such as the H-He massive atmospheres possessed by sufficiently massive planets), or heavier gases in atmospheres of lower mass planets, could be less immune. For example, Lammer et al. (2011) show that an Earth-size planet with a nitrogen atmosphere may be affected by a radiation level only 10 times the solar X and UV radiation, if present for a sufficiently long time, such as the early evolutionary period of low mass RDs. Therefore, unless rebuilt during the later, more stable stages of the host star, such RDPs may suffer from a thinned atmosphere.

## 6. Liquid water and tidal locking within Habitable Zones of RD stars

The Habitable Zone is defined as the region around a star in which the equilibrium surface temperature of a planet with an atmosphere similar to that of Earth could support liquid water at least on part of the surface, at least part of the time. This is a *sine qua non* for life (e.g. Kasting *et al*, 1993) – but always with the caveat of "life as we know it". In the Solar System the HZ extends approximately between the orbits of Venus and Mars, with Earth approximately at its center. Planets with a thicker atmosphere would be able to have surface liquid water further out from the Sun, because of the greenhouse effect. Planets with a significantly higher albedo could support liquid water on orbits nearer to the Sun. It should be noted however, that liquid water has been detected in other places in the Solar system, on planet satellites (e.g. Europa and Enceladus) far outside the HZ (Kangel, *et al.*, 2000).

Life on RDPs that evolves under water (as did life on Earth for the first 3 billion years) would be protected from flares and XUV radiation, as those would penetrate only the thin upper layer of water. However, even RDP oceans are not immune to evaporation by the radiation from the host RD during its extended pre-Main Sequence phase, causing a runaway greenhouse effect. During this early runaway phase, photolysis of water from RDPs in the habitable zone (for Earth-size and smaller planets) could result in hydrogen/oxygen escape to space of large quantities of water, up to several times that of all the Earth's oceans (Tian et al. 2014; Luger and Barnes 2015). This does not necessarily preclude the existence of surface water on all RDPs during later stages. First, the extended runaway scenario may be less important for RDPs at the outer HZ and for higher-mass RDs. Second, water may be required by accretion of cometary and asteroidal water after the runaway period, if the latter is shorter than the period of bombardment by comets and icy asteroids. Third, in the case of massive super-Earth RDPs, water may be reproduced from the oxygen and hydrogen retained in the atmosphere after early-on splitting of water.

As noted above, the HZ for planets orbiting RD stars would be much closer to their host star than for solar-type stars. This proximity would probably result in tidal locking, with only one hemisphere continuously exposed to the star, much as our Moon is tidally locked to Earth (Gladman *et al.*, 1996). It was originally thought that tidally locked planets, having one side continuously illuminated, with the other hemisphere in the dark, could not be hospitable for life, as their dark side would be permanently frozen and their atmospheres would be too turbulent, as a result of planet-wide winds of hurricane proportions. However, recent calculations indicate that even at the light-dark terminator, winds would be mild, especially in the presence of planet oceans, which mediate heat transport (Hu and Yang, 2014). Moreover, the drag effect of a massive atmosphere like that of Venus could counteract the tidal locking, and recent calculations show that even a relatively thin atmosphere can drive terrestrial planets' rotation away from synchronicity (Laconte et al. 2015). Other consequences of tidal locking are discussed below.

# 7. The Importance of Oxygenic Photosynthesis for the evolution of Complex Life.

Oxygen is a highly active chemical element, hence its presence in an atmosphere, in more than traces, indicates that it is being continuously produced. On Earth it is a result of the activity of plant OP, especially oceanic cyanobacteria and algae. Cyanobacteria were the first water splitting photosynthetic organisms on Earth. Consequently, Oxygen is a major bio-signature in the search for life outside of Earth. However, as noted above, significant amounts of oxygen could be produced on young RDPs abiotically (e.g. by near UV radiation splitting water molecules). This abiotic oxygen could also produce a false biosignature (Tian 2014; Luger and Barnes 2015). Earth's oxygen-rich atmosphere has been both good and bad for life in general and for plants in particular (Gale, 2009). One of the early effects of OP was inducing Ice Ages, although their driver was not only the increase of oxygen; also the concomitant reduction of atmospheric $CO_2$ and $CH_4$, which absorb near infrared radiation, contributed to planet cooling. The evolution of Earth's oxygen rich atmosphere is depicted in Fig.1.

Another negative effect of the oxygen-rich atmosphere on Earth life is that it inhibits nitrogenase, the enzyme that certain anaerobic bacteria use to fix $N_2$. Although nitrogen is required by all life forms and constitutes some 80% of Earth's atmosphere, this is the only known process by which advanced life can (indirectly) acquire nitrogen from the atmosphere (Raymond $et$ $al$, 2003).

Today, the OP mechanism (especially in higher plants) is itself partly inhibited in its $CO_2$ fixing activity by the very oxygen it produces. The primary plant enzyme that captures $CO_2$ is Ribulose Bi-Phosphate Carboxylase (RUBISCO). At today's atmospheric level of $O_2$ (21% v.v.) the oxygen competes with $CO_2$ (only 0.04% v.v.) for the fixation site. This is true for 80% of the world's vegetation. Only in the very recent 150 million years have new higher plants evolved (of the so-called C4 type) which can overcome this effect. In the majority of C3 plants, oxygen reduces $CO_2$ fixation by some 25% (see e.g. Lambers $et$ $al.$, 1998 and Sage 2003, for a full discussion).

As noted, the high concentration of $O_2$ in the atmosphere provides protection from ultraviolet radiation (<300 nm). Without it, the unfiltered UV reaching Earth is energetic enough to cause damaging ionization in bacteria, plant and animal cells. The oxygen atmosphere absorbs some of the UV and in doing so produces ozone, which strongly absorbs UV. This protective atmosphere appeared on early Earth shortly following the evolution of the Cyanobacteria, the first plants which released oxygen from water (Ward et al, 2015). It has been called the GOE, Great Oxygen Event (Fig. 1). It enabled life to emerge from the protection of the oceans and shaded niches and invade dry land (Wayne, 1992, Kasting and Catling, 2003).

The high oxygen atmosphere also enabled the evolution of aerobic respiration, which releases nine times as much energy (per molecule of carbohydrate or other substrate) than produced by $anaerobic$ respiration (e.g. Lehninger, 2005). A rapid supply of energy is a primary requirement of mechanically active, complex, multi-cellular life. Advanced nervous systems are also heavy expenders of energy. The downside for plants was that this also enabled the evolution of some of the strongest competitors of the Earth's vegetation such as insects and animals. *Homo sapiens sapiens* is only the latest scourge. Together with the UV protection of the high $O_2$ atmosphere, aerobic respiration is considered to have been a major factor that drove the Cambrian "explosion" (Catling *et al*, 2005). This was a period from ~ 550 to 470 million years ago when life advanced from 3.3 billion years of mainly single celled organisms to multi-cellular, complex life forms (Fig.1). For all these reasons, OP and conditions for OP are primary targets in the search for advanced extrasolar life (Cockell *et al*, 2009). Although there were (and still are), a number of problems for the biota resulting from the oxygen-rich atmosphere produced by OP, without its appearance on Earth there would probably have been no complex life as we know it.

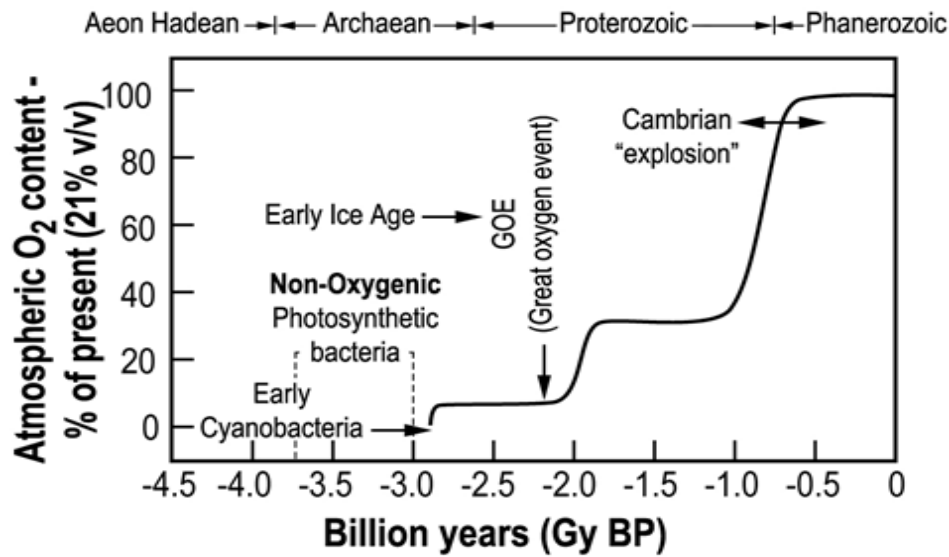

**Figure 1.** Time course of the development of Earth's oxygen rich atmosphere as a percentage of the present oxygen content (21% volume/volume).

## 8. The Environment on RD planets in relation to Oxygenic Photosynthesis and Plant Growth Patterns on Earth

Our Sun is a G-type star, with a surface temperature of 5,860 K. At this temperature, the solar radiation spectrum incident on Earth after passing through the atmosphere extends from ~250 - 2000 nm. Chlorophyll, the main active pigment of OP in nearly all of Earth's vegetation, has evolved to utilize only about 48% of this radiation, in the waveband extending from 400 nm to 700 nm, where its absorbance sharply drops off. This is called the Photosynthetically Active Radiation (PAR) waveband (Fig. 2).

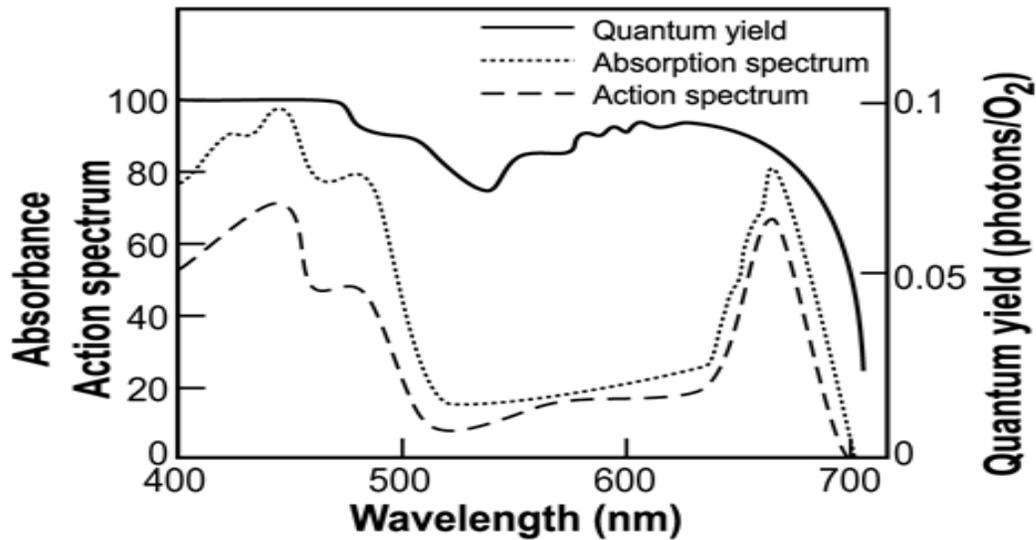

**Figure 2.** Absorption, action and quantum yield spectra of chlorophyll during oxygenic photosynthesis, showing the sharp cut-off at 700 nm. (Data collated from different sources).

RDs are late K and M-type stars. Their surface temperatures are significantly lower than that of G stars, such as our Sun. Consequently, the spectrum of their radiation is shifted towards longer wavelengths, in the near infrared (NIR). In view of the sharp chlorophyll absorption cut off at 700 nm (Fig. 2) this was first thought to be a precluding factor for OP on RD star planets. However, a considerable part of the radiation would still be in the PAR waveband (Fig.3). This, together with the continuous illumination on the tidally locked RD planets, could well provide sufficient energy for OP and growth (Figs. 3, 5 and 6).

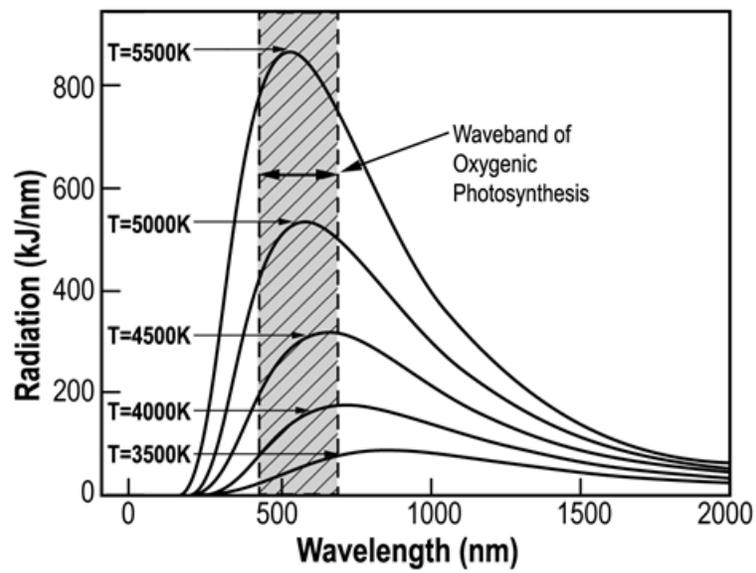

**Figure 3.** Black body Planck curves of different temperatures, showing the effect of the shifting of the radiation of RD stars towards the NIR, on the energy available in the oxygenic photosynthesis waveband. (All stars have very nearly black body spectra).

Contrary to some early and oft quoted reports (e.g. Emerson and Lewis, 1943) the OP drop is not caused by lack of energy in photons of wavelength >700 nm. The theoretical energy requirement for splitting a water molecule is 1.23 eV (albeit in reversible systems at standard temperature and pressure and 0 pH) while the quantum energy of a photon at 700 nm is 1.77 eV. The quantum energy of the photon only drops below 1.23 eV at wavelengths >1000 nm. It has also been pointed out that under certain circumstances water splitting may be possible at wavelengths >700 nm in stepped photosynthetic systems (Kiang *et al.,* 2007a, b) but at a cost of lower overall efficiency (Wolstencroft and Raven, 2002). Ordinary splitting of water actually occurs at 680 nm. Part of the quantum energy of the 680-nm photon is also used to drive other electron transfer reactions of photosynthesis. Moreover, some cyanobacteria do carry out OP at wavelengths >700 nm; e.g. *Synechococcus leopoliensis* has a peak absorbance at 670 nm and uses Chlorophyl a, and *Acaryochloris marina* has a unique pigment (Chorophyll d) with peak absorbance at 710 nm (Mielke *et al.,* 2011).

The possibility of OP at wavelengths >700 nm raises the question as to why, in 3 billion years of photosynthesis evolution, most Earth plants have not evolved to utilize more than the 400-700 nm waveband, which contains only about 48% of the available solar electromagnetic radiation. This question is very relevant to RDPs. A first possible explanation is that plants and OP evolved in an environment lacking NIR radiation. A layer of ~15 cm of pure water absorbs almost all NIR >700 nm. The primeval oceans, beneath an atmosphere with little oxygen, did not consist of pure water but also contained other species including, for example, dissolved, reduced $Fe^{++}$ salts, which also strongly absorb NIR (Curcio and Petty, 1951), increasing the absorption even more.

A second explanation is that on dry land lack of radiation energy is rarely a major factor limiting OP and growth of plants from germination to seed. Moreover, Milo (2009), calculated that OP at around 700 nm is the most theoreticaly efficient in its use of quantum energy. The present OP mechanism expends ~12 photons in the fixation of each molecule of $CO_2$; OP at longer wavelengths would require more (Raven, 2007). Consequently, there would have been no environmental evolutionary pressure to utilize radiation >700 nm. When plants emerged from sea to land, not only did they not evolve mechanisms to utilize the now available NIR, but actually rejected it. Plant leaves reflect and do not absorb a large part of the incident NIR (Fig. 4).

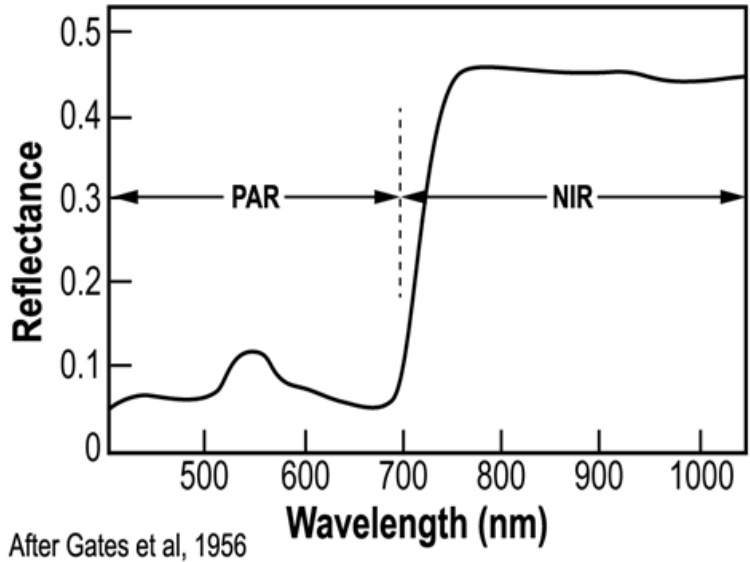

**Figure 4**. Reflectance of a typical plant leaf in the visible and near infrared (after Gates *et al.*, 1956).

As seen in Fig.4, some 45% of the 700-2000 nm waveband, which contains >40% of the solar electromagnetic radiation energy incident on the planet surface, is reflected, while in many plant species a further portion of the NIR is transmitted. Evidently reduction of leaf NIR absorbance, and hence of heating, transpiration and water loss, was more advantageous in the evolution of land plants than extra photosynthesis.

As RDPs in HZs would probably be tidally locked, only plants on the star-facing hemisphere would be able to carry out OP. There is a however a clear advantage for such plants as they would be exposed to continuous radiation. The situation is analogous to plants growing on Earth, in the summer months, in very northerly (and southerly) latitudes. In such regions, vegetation is often lush, despite there being only a very short growing season, bordered by long periods of low light and extreme cold. The reason for this burst of growth is that during the short growing season solar radiation is almost ideal for OP, with moderate light intensities for almost 24 hours a day. This contrasts to conditions in the concomitant summer periods at lower latitudes, where for much of the day the radiation intensity exceeds that which can be utilized for OP. This is shown schematically in Fig. 5.

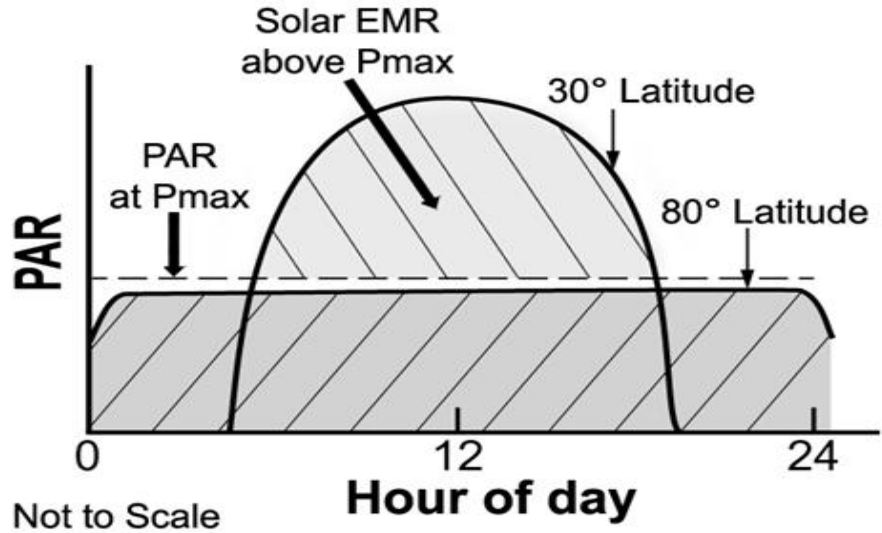

**Figure 5.** Photosynthetically Active Radiation (PAR) during the "Grand Period of Growth" incident on northern versus mid-latitude plants, on Earth – an analog for plants growing on tidally locked Red Dwarf star planets.

The super-optimum radiation ($>P_{max}$) during much of the day in low latitude ecosystems (Fig. 5) may even be damaging to the OP mechanism itself and cause photo-oxidation. Before the onset of photo-oxidation, $P_{max}$ is mainly determined by the capacity of the photosynthesis system (Lambers *et al.*, 1998). Photo-oxidation certainly leads to an increase in water loss, which could reduce OP. This happens initially by causing leaf stomata to close, which interferes with the diffusion of $CO_2$ into the leaf, and then, eventually, by damaging the photosynthetic apparatus. Excessive plant incident radiation would be absent on RDPs as the vegetation on such tidally locked planets would distribute itself to those regions having the radiation intensity most adapted to its species requirements (Heath *et al.*, 1999, and Fig. 6).

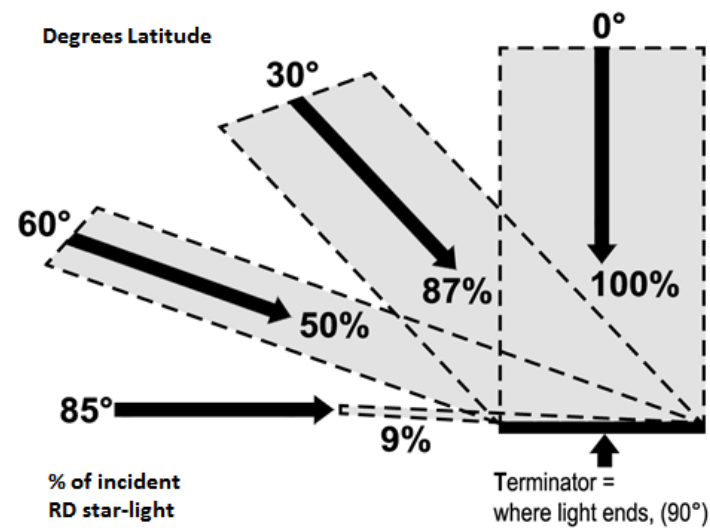

**Figure 6.** The distribution of radiation on the surface of a planet tidally locked to its star. The sub-stellar point corresponds to $0^o$.

As depicted in Fig. 6, radiation on the curved, star-facing surface of the planet, is reduced in accordance with the angle. It would decrease from a maximum, in the region immediately facing the star, to zero at the terminator.

## 9. Red Dwarfs and the abundance of biotic life

We now demonstrate the effect of adding RDPs to the population of potential biotic worlds on the probability of detecting extra-terrestrial life. Consider the distance to our nearest biotic neighbor planets, given by (Wandel, 2015)

$$d_b \sim 10 \, (R_b \, F_b)^{-1/3} \text{ light years} \qquad \text{(eq. 2)}$$

Here $Fb$ is the probability of biotic life appearing on an Earth-sized planet in the HZ, and $R_b$ is the rate at which habitable planets are formed in the Galaxy

$$R_b = R^* \, F_s \, F_p \, F_e \, n_{hz}. \qquad \text{(eq. 3)}$$

In eq. 3 there are two astronomical factors: $R^*$ - the rate of star birth in the Milky Way (~10 per year) and $F_s$ - the fraction of stars suitable for evolution of life; both are well established. The three other "planetary" factors, $F_p$, $F_e$ and $n_{hz}$ were totally unknown up to a few years ago, and could be reasonably well estimated only after the Kepler results. $F_p$ is the fraction of stars that have planets, $F_e$ the fraction of Earth-size planets, and $n_{hz}$ is the number of such planets per star, within the Habitable Zone.

As discussed in the introduction, Kepler and other recent observations have shown that most, if not all stars have planetary systems, so $F_p \sim 1$.

If only planets around solar-type stars are considered, then the fraction of stars suitable for evolution of life is the well-known fraction of solar-type stars in the Milky Way, $F_s \sim 0.1$ (~8% of all Main Sequence stars are G-type). As discussed above, according to recent analyses of the Kepler data only 2-4% of all solar-type stars have small, Earth-sized planets orbiting within their HZ (Foreman-Mackey, Hogg and Morton, 2014; Farr, Mandel and Stroud, 2014), hence for solar-type stars $F_s F_e n_{hz} \sim 0.002$-$0.004$. If in addition one excludes planets around stras in multiple stellar systems (70% or more of all individual stars, in particular RDs, see section 3), this figure could be smaller by a factor of 4-5, which gives the most severe constraint on the rate at which habitable planets are formed in the Galaxy, $R_b \sim 10 \times 0.2 \times 0.002 = 0.004$ per year.

However, if in addition RD stars (single or in multiple systems) may host habitable planets, then $F_s \sim 0.9$. According to the analyses of Dressing and Charbonneau (2015) and Kopparapu (2013) about half of the RD stars have habitable planets, which leads to much larger figures, $F_e n_{hz} \sim 0.4$ and $R_b \sim 4$ planets per year.

Depending on the assumptions as to which stellar type is suitable for life, the rate $R_b$ could thus vary by a factor of 1000, in the range, ~0.004 - 4. Using Eq. 3 this would yield a factor of 10 in the estimate of the distance to our nearest biotic neighbor planets. This is shown in Fig. 7, which shows the distance given by Eq. 2 as a function of the biotic factor $F_b$, for three choices of host-stars. Fig 7 also provides planet-statistics, as discussed above: (a) solar type, single stars with a low fraction of Earthsize HZ planets, ($R_b=0.004$), (b) solar type, single or multiple stars with a high fraction of Earthsize HZ planets, ($R_b=0.1$) and (c) single or multiple Red Dwarfs ($R_b=4$). It is easy to see that assuming life can evolve on RDPs significantly enhances the abundance of potential biotic planets. For example, if $Fb = 0.1$ the estimate of the distance to our nearest biotic neighbor would vary between 140 light years in the conservative single solar-type star case and fourteen light years in the multiple RD case.

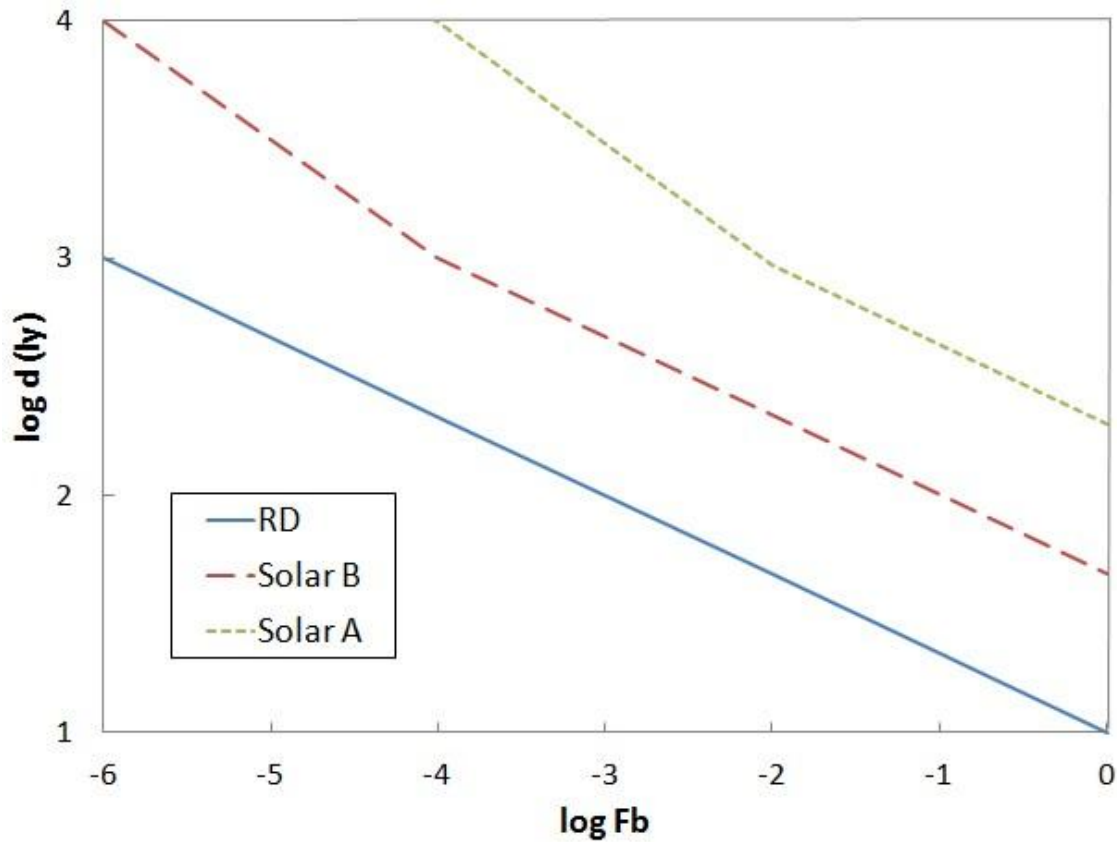

**Figure 7.** The distance to our nearest biotic neighbors as a function of the biotic factor $F_b$, for several choices of host star types and Earthlike planet fractions. The upper, short-dashed curve labeled "solar A" is for the most conservative case – only single, solar type host stars, assuming that 2% of them have Earthsize planets in the HZ. . The middle long-dashed curve labeled "Solar B" is an intermediate case – solar type host stars in single or multiple stellar systems, assuming

that 20% of them have Earth-size planets in the HZ. The lower solid curve marked RD depicts single or multiple RD host stars assuming that 50% of them have Earth-size planets in the HZ.

As a concrete example we assume the probability of biotic life, $F_b$, to be between 0.001-1. Then, according to Fig. 7, the distance to the nearest biotic extrasolar planet could vary in the range of 10-100 light years in the RD case, and 100-1000 ly in the conservative solar-type host star case. In the near future the probability $F_b$ may be estimated observationally by spectral analyses of the light reflected from exoplanets (e.g. Demming et al. 2009; Loeb and Maoz 2013), using space telescopes such as JWST or Starshade.

# 10. Conclusions

Adding RDPs to the potentially life supporting planets increases the probability of finding biotic planets by a factor of up to a thousand, and reduces the distance to our nearest biotic neighbor by up to 10. Recent works have demonstrated that the Habitable Zones of RD stars may be less hostile to life than thought previously. For all the reasons given above, Oxygenic Photosynthesis and perhaps complex life on planets orbiting Red Dwarf stars may be possible. Moreover, the vast number of RDPs in the galaxy seems to make this possibility probable, although to date we still have only the single example of life on Earth. It has been argued that photosynthetic pathways may develop to exploit RD radiation at wavelengths between 700-1000 nm, despite lower overall photosynthetic efficiency. However, we suggest here that the evolutionary pressure to do so on RDPs would be small, as tidal locking will provide continuous illumination with a considerable component of photosynthetic 400-700 nm radiation. This is analogous to the lush summer growth of vegetation on Earth, at very high latitudes. Even so, any life that does develop would not be the same as on Earth. For example, there would be no circadian and seasonal rhythms in the flora and fauna, nor would complex life forms have daily sleep patterns. Moreover, extra-terrestrial life may develop using a genetic code different from the single example of life that appeared on Earth - with unpredictable consequences.
The James Webb, Giant Magellan and Thirty Meter telescopes, planned to be operational in the next decade, may provide evidence of the life predicted here. This could be obtained by the detection of biomarkers in the atmosphere of planets transiting compact stars. Oxygen produced by OP could be such a bio-signature, although remnants of abiotic $O_2$ may confound the results.